\documentclass[11pt,a4paper]{article}

\usepackage[utf8]{inputenc}
\usepackage[T1]{fontenc}
\usepackage{amsmath}
\usepackage{graphicx}
\usepackage{booktabs}
\usepackage{subcaption}
\usepackage{ragged2e}
\usepackage{verbatim}
\usepackage{xcolor}
\usepackage{url}
\usepackage{comment}

\usepackage[round,comma]{natbib}

\usepackage[a4paper,width=160mm,top=35mm,bottom=30mm]{geometry}
\usepackage[colorlinks=true,linkcolor=black,urlcolor=black,citecolor=black]{hyperref}

\title{A Theory-Based AI Automation Exposure Index: \\
Applying Moravec's Paradox to the US Labor Market}
\author{Jacob Schaal\\
London School of Economics\\
\\
Mentor: Herbie Bradley\\
Research Manager: Nandini Shiralkar}
\date{}

\begin{document}

 
\begin{titlepage}
\begin{center}
    
\LARGE
Cambridge ERA AI Governance Research Fellowship\\

\vspace{0.5cm}
      
\rule{\textwidth}{1.5pt}
\LARGE
\textbf{A theory-based AI automation exposure index: \\
Applying Moravec’s Paradox to the US labor market }
\rule{\textwidth}{1.5pt}
   
\vspace{0.5cm}
      
\large
Mentor: Herbie Bradley \\
Research Manager: Nandini Shiralkar\\

\vfill

\Large
\textbf{Jacob Schaal}
 \\
\large
London School of Economics\\

\vfill
\textbf{Abstract:}
\\
This paper develops a theory-driven automation exposure index based on Moravec's Paradox. Scoring 19,000 O*NET tasks on performance variance, tacit knowledge, data abundance, and algorithmic gaps reveals that management, STEM, and sciences occupations show  the highest exposure. In contrast, maintenance, agriculture, and construction show the lowest. The positive relationship between wages and exposure challenges the notion of skill-biased technological change if AI substitutes for workers. At the same time, tacit knowledge exhibits a positive relationship with wages consistent with seniority-biased technological change. This index identifies fundamental automatability rather than current capabilities, while also validating the AI annotation method pioneered by \citet{eloundouGPTsAreGPTs2024} with a correlation of 0.72. The non-positive relationship with pre-LLM indices suggests a paradigm shift in automation patterns. 

      
\vfill


\vfill

\end{center}
\end{titlepage}


\pagenumbering{arabic}
    
\tableofcontents

\newpage
\pagenumbering{arabic}
\section{Introduction}
\label{intro}

%

The labor market effects of artificial intelligence (AI) have entered mainstream discussion recently: For instance, the Financial Times asked in a headline: "Is AI killing graduate jobs?" \citep{Murray2025AIJobs} On the other hand, automation has affected the labor market for decades, and AI remains particularly weak in some domains that are easy for humans, which is called Moravec's Paradox \citep{moravec1988mind}.
\\
I introduce a theory-based index using  this paradox. Evolutionary optimization could explain this (Erdil, 2024), as functions that are perceived as difficult for humans have high variance in human performance, likely because they were developed late in evolution, such as mathematics or coding. For instance, AlphaGo beat the world's best Go Player Lee Sedol in 2016 which is a highly strategic and complex board game.   On the other hand, physical and somatosensory skills exhibit low variance among (non-disabled) people. Additional theoretical reasons are data abundance, tacit knowledge, and the algorithmic efficiency gap. \footnote{All annotated task-level and occupation-level exposure scores are available at \href{https://bit.ly/4pguKNd}{https://bit.ly/4pguKNd}}

First, tasks with more digital training data are more manageable for AIs to perform. Second, some tasks require a lot of tacit knowledge, which is sometimes referred to as Polanyi's Paradox \citep{autorPolanyisParadoxShape2014}. The algorithmic efficiency gap relates to the ways in which the brain has naturally evolved for specific tasks, in contrast to neural networks that have a different architecture, such as visual segmentation or sensorimotor control. 
Current AI automation indices disagree by over 70\%.\cite{Webb2019}'s patents correlate only 0.31 with Felten's benchmarks \citep{Felten2021}. 
My theory-based approach reveals a striking pattern: management and STEM occupations exhibit the highest exposure to automation. At the same time, maintenance and construction workers face the lowest risk, suggesting that AI may reduce rather than exacerbate wage inequality. This is consistent with the majority of current empirical studies (e.g. \citet{Noy2023}, at least within occupation).
In general, existing AI automation exposure indices rely on current capabilities, whether they're task feature analyses, patent mapping, or expert surveys, and don't address general automatizability. This limits the usefulness of these indices. Older indices are hardly related to newer ones (e.g., post-LLM), which raises doubts about the entire research area. 

My contribution includes two substantive findings and one methodological validation. First, I present a theory-based automation index, rather than one based on current technological capabilities, which offers predictions that may remain stable as AI advances. Second, I demonstrate that tacit knowledge, which exhibits a unique positive relationship with wages while being negatively correlated with my other theoretical dimensions, aligns with recent evidence of seniority-biased technological change, where early-career workers face disproportionate displacement. 
Methodologically, I validate that the AI annotation approach is remarkably robust. Despite using a completely different prompt than Eloundou et al. (2024), I achieve a correlation of 0.72 with their results, and inter-model correlations exceed 0.8 for reasoning models. This demonstrates the robustness of the AI annotation method.

This working paper proceeds as follows:
\begin{itemize}
    \item Chapter \ref{Literature Overview} reviews the literature
    \item Chapter \ref{Outlook} overviews recent literature on the effects of generative AI, especially on early-career workers, followed by a brief theoretical outlook
    \item Chapter \ref{methodology} explains the theory and empirical methodology
    \item Chapter \ref{results} presents the results
    \item Chapter \ref{Discussion} discusses the approach and the results
    \item Chapter \ref{conclusion} concludes
\end{itemize}


\section{Task-Based Frameworks, Automation, and AI: A Review}
\label{Literature Overview}
The relationship between artificial intelligence and unemployment has become a heated topic in economic discourse. AI, as well as other forms of automation, both destroy and create jobs. This overview examines how various methodologies attempt to understand AI's potentially transformative role.

\subsection{Methodologies}
The literature reveals three broad methodologies to predict or explain automation: Task patent mapping (e.g., \cite{Webb2019}), --- at least for predictions--- automation forecasting surveys (e.g., \cite{Gruetzemacher2020}), and task feature analysis \citep{Autor2013}. I harness the latter approach since it is the most common and empirically sound in the literature. It allows for fine-grained analysis of the specific activities a worker performs, typically in a job.
One of the core challenges for the two task-based methods is the difficulty in capturing the nuanced interplay between AI, task structure, and the human workforce. AI's impact on the labor market is not uniform; it varies across industries, job types, and tasks. Task-based approaches do not carefully distinguish between the full and partial automation within jobs and the deskilling of specific jobs due to a shift towards "less-skilled" tasks.


\cite{Webb2019} introduces an innovative approach, employing text analysis to measure the exposure of job tasks to automation technologies. By comparing job task descriptions with patent texts, Webb's method can predict which occupations are more likely to be affected by new technologies, such as AI. This approach directly links technological advancements with their potential impacts on specific job tasks. AI exposure is also strongly correlated with the average wage in a particular occupation. Webb's findings suggest that, unlike previous automation technologies, AI primarily targets high-skilled tasks, which may reduce wage inequality, except for the highest earners. This seems prescientific, given the latest evidence discussed later about generative AI (e.g., \cite{Noy2023}).
\\
In contrast, \cite{Gruetzemacher2020} presents a survey-based study focusing on AI's potential for extreme labor displacement. The survey, conducted among AI conference attendees in 2018, reveals that practitioners expect a significant portion of current human tasks to be automated by AI in the near and mid-term. The median prediction suggests that  60\% of tasks are completed within ten years. More strikingly, there's a forecasted 50\% probability of AI systems being able to automate 90\% of human tasks within 25 years and 99\% within 50 years. The study seems relevant for more long-term and speculative predictions about job losses due to AI. Still, a consensus among computer scientists only tells so much about the labor market. I prefer an approach immediately grounded in labor market data, which does not require faith in expert predictions.
\\
Therefore,  I take a more conceptual and standard route with the task feature analysis, which is described by \cite{Autor2013}. This approach shifts the focus to the specific tasks that labor and capital perform. Autor argues that changes in allocating these tasks, influenced by technological advancements and globalization, have led to a restructuring of labor demand. This restructuring, which took the form of labor market polarization over the last decades, manifests as increased demand for high-skill and low-skill jobs coupled with a decline in middle-skill jobs  \citep{Autor2013b}. The task approach provides a framework to understand how AI and automation may reallocate tasks, reshaping job structures and wage patterns. AI annotation, in particular, is a valuable method for systematically analyzing the features of a wide range of tasks. 

A recent method for analyzing the effect of AI on tasks is AI annotation, as pioneered by Eloundou et al. (2024). They annotate the O*NET task database via GPT-4 and human annotators to get automation scores for over 19,000 tasks. Their preferred exposure metric measures whether the adoption of GPT-4 would potentially lead to a 50\% time reduction. 

GPT-4 and humans rate the skills of large language models (LLMs) as surprisingly similar, illustrating the expansion of tasks that AI can achieve. According to the working paper, 80\% of the US workforce will be affected by AI in their work life, with 19\% experiencing that at least half of their functions are impacted. 

The relevance of this exposure index for AI usage has been confirmed using large-scale Bing Copilot data, with correlation factors exceeding 0.7 \citep{tomlinsonWorkingAIMeasuring2025}.

Their approach inspires mine and follows their methodology, but is theoretically grounded in multiple theories and doesn't depend on current capabilities. On the other hand, it doesn't follow for precise categorization based on time savings or tool usage. 
\\
The methodologies employed by \cite{Webb2019}, \cite{Gruetzemacher2020}, and \cite{Autor2013} illustrate the multifaceted challenges in predicting and understanding AI's impact on the labor market. These studies highlight the need for nuanced, context-specific approaches to understanding the complex dynamics of AI-driven labor market transformations. These methods differ significantly in which jobs are most exposed to AI. The ongoing debate on AI's role in job displacement versus job creation continues to evolve, reflecting the rapid advancements in AI technologies and their diverse applications in the labor market. As I will discuss later, labor economists' methods are also in danger of becoming obsolete by AI progress \citep{Felten2023}.
\subsection{Evidence of Labor Market Effects of automation exposure}
 \cite{Acemoglu2020} finds adverse effects of robots on wages and employment. \cite{Brynjolfsson2018} analyzes the labor market impact of machine learning, while \cite{Restrepo2023a} comprehensively surveys the existing automation literature.  \cite{autorApplyingAIRebuild2024}hypothesizes that AI differs from robots and that this might cause wage polarization to reverse. 
\\
\cite{Acemoglu2020} examine the impact of industrial robots on US labor markets. It theorizes that robots may decrease employment and wages, with their local impacts being estimable through exposure variation to robots, defined by industry-level advances in robotics and local industry employment. The study finds robust adverse effects of robots on employment and wages across commuting zones. It differentiates the impact of robots from other capital and technologies, showing that one robot per thousand workers reduces the employment-to-population ratio by 0.2 percentage points and wages by 0.42\%.
Unlike prior research that often focused on task automation, Acemoglu and Restrepo provide a nuanced analysis of job automation. However,  the potential difference in the nature of robots (tangible, rivalrous assets) compared to AI algorithms (intangible, non-rivalrous) remains a limitation of the external validity. Therefore, AI might affect wages and unemployment differently from robots. This distinction is crucial in avoiding false analogies about the effects of  AI on the labor market.

\cite{Brynjolfsson2018} comprehensively analyzes tasks affected by machine learning (ML), closely related to AI. The study assesses the "Suitability for Machine Learning" (SML) of various functions in the O*NET database, which contains occupational skills from the US Department of Labor, providing the background for my analysis. In contrast, earlier research relied on expert predictions regarding the automatizability of tasks. They reveal that ML impacts different occupations than earlier waves of automation, with most jobs comprising at least some SML tasks. The key findings include that few occupations are fully automatable using ML, and realizing ML's potential usually requires redesigning job task content.

\cite{Restrepo2023a} comprehensively surveys existing literature on the impact of automation on labor markets. The paper compiles and synthesizes diverse research findings, shedding light on the multifaceted nature of automation's impact. Restrepo pays particular attention to task-based models and their implications in understanding how automation influences job displacement. This review stands out for its breadth, covering various methodologies and empirical evidence that collectively depict a nuanced picture of automation's role in reshaping labor dynamics. The paper highlights the complexities involved in evaluating the impact of automation and AI on employment, wages, and task allocation within the labor market. He concludes with a section on AI by separating generative AI from other technologies and calling for future research on the developing topic.

Autor (2024) hypothesizes that AI could reverse wage polarization in recent decades. It enhances the expertise of ordinary workers and enables a broader segment of the labor market to process information for critical decisions. Higher AI exposure for high-wage jobs is in line with this hypothesis.
Reflecting on these insightful works, it is clear that the AI revolution presents a paradox. On one hand, there is an undeniable potential for AI to catalyze unprecedented productivity growth \citep{trammellEconomicGrowthTransformative2023}. On the other hand, this technological leap forward presents significant challenges, including labor displacement and widening inequality. Collectively, these papers highlight the uniqueness and transformative power of AI, the importance of focusing on technology-specific exposure indices, and the need for ongoing observation of the labor market by policymakers. 
As I will show now for generative AI, the distributional effects of innovation shift rapidly depending on the underlying technology. 


\section{Outlook Toward Generative AI and Early-Career Unemployment}
\label{Outlook}
\subsection{Recent Research on the Labor Market Effects of Generative AI}
Following the release of ChatGPT in November 2022, this form of AI has impacted the labor market in a manner that prior automation indices had hardly predicted. To gain an accurate understanding of AI's current and potential future effects on wages and unemployment, examining the most up-to-date research is indispensable. Some experiments empirically evaluate AI's short-term effects and mostly find more negative effects for higher-educated workers (\cite{Noy2023}; \cite{Hui2023}). The last few years after the pandemic have been characterized by a hot labor market, drastically reducing wage inequalities \citep{Autor2023}. A cautious note that encapsulates the early stage of the literature comes from \cite{Otis2023}, whose randomized controlled trial showed that a GPT4-powered chatbot increased the wage differential between Kenyan entrepreneurs. 
\\
\cite{Noy2023} expose their experiment participants to writing tasks specified for their jobs. They measure the time taken, output quality, and job satisfaction. The treatment group has the opportunity to revise their writing based on feedback from ChatGPT.  This raises productivity by reducing the necessary time by 0.8 standard deviations, while improving the quality of the text by 0.4 standard deviations. These are rather large effect sizes, but they are only measured on specific tasks and have only short-term effects. These findings are consistent with \cite{Hui2023}, who find that freelancers' quality mediates the impact of AI, with higher-skilled freelancers on the Upwork platform being more affected.
Nonetheless, they found that the introduction of ChatGPT in November 2022 had harmful effects on the wages and employment of freelancers. Workers in the most affected occupations experience highly significant reductions of 2\% in orders and 5\% in wages. Whether these effects persist over the long term and how AI impacts the overall welfare of all stakeholders remains an interesting area for future research. 
Since the annotators do not stem from diverse backgrounds of workers, subjectivity and limited experience curb the quality of the labeling. How these changes in task performance alter various industries and nations remains equally relevant for future research in analyzing the limitations of LLMs for labor. Nonetheless, the perspective of AI as a general-purpose technology remains convincing. 
\\
This technology impacts a tight labor market: Labor is highly sought after, which led to solid wage rises since the pandemic, especially for low-skilled workers \citep{Autor2023}. Consequently, the rise in wage inequality over the last four decades could be reversed by a third over three years. Although this wasn't the result of AI, the first results consistently hint at a future decline in the college wage premium. A good indicator that the state of the labor market causes wage increases is the concentration of these increases among job-changing workers, or technically speaking, among workers with high labor supply elasticity. Therefore, the strong labor market gave workers more bargaining power, which they utilized by increasing their quit elasticity. In general, the labor market is in the hottest state in decades, underlining the fortunate timing of AI \citep{Autor2023}. On the other hand, wage increases also provide more substantial incentives for more automation. The focus of the following subsection will be how this will play out over the longer term. 
\\
A recent field experiment by \cite{Otis2023} investigates the effects of a GPT-4-powered business mentor via WhatsApp for Kenyan entrepreneurs. Though not immediately relevant to the labor market, it challenges conceptions of AI-induced wage compression. The study proxies wages by observed productivity, in the form of revenues and profits of entrepreneurs of varying skill levels. In contrast to the literature previously discussed, they estimate an adverse effect of 0.11 standard deviations for low-skilled entrepreneurs, while high performers improve by 0.20 standard deviations. Therefore, the income gap increases contrary to the findings of other research. This might be because low-performing entrepreneurs focus on more challenging business tasks, indicating that they are negatively selected into these functions. Therefore, this is not inconsistent with the prior-discussed literature, which demonstrates that low performers gain more from AI at a given task. High-skilled entrepreneurs may also possess the necessary tacit knowledge to utilize AI effectively.

\subsection{Recent Research on Early Career Unemployment}
 \citet{brynjolfssonCanariesCoalMine} use Eloundou's AI annotation-based index to show that early-career workers in the most exposed occupations have experienced a 13\% relative decline in employment. This decline is particularly pronounced in occupations where AI automates rather than augments human labor. They harness Anthropic's Economic Index \citep{handaWhichEconomicTasks2025} to classify the jobs. They suggest that tacit knowledge might explain the advantage of more experienced workers in a labor market impacted by AI. I will test tacit knowledge as one factor of my exposure index. 

\citet{lichtingerGenerativeAISeniorityBiased2025} validates this result. They  utilize large-scale job posting data to classify firms as AI adopters, while also employing Eloundou's pioneering index. Those AI-adopting firms started 2023 to reduce their early-career hiring by 22\%. They call it seniority-biased technological change, which, in contrast to skill-biased technological change, doesn't drive a wedge between workers with high and low education but between senior and junior workers due to tacit knowledge.

\subsection{Theoretical-Based Outlook Toward the Future of Work}

\cite{Korinek2022} explore the consequences for workers of an advanced AI that substitutes for human labor. Therefore, labor might become abundant if machines can perform every task more efficiently than humans. This could drastically curtail labor demand, leading to falling wages. Under this scenario, the authors advocate for phasing out labor, starting with less productive workers, and introducing a universal basic income based on the principles of a utilitarian social planner. This is only one, and a rather pessimistic, prediction about the future of work. Empirical research remains of utmost importance to improve economic predictions. One instance is the following analysis of what I believe is the first theory-based automation exposure index.


\section{Methodology}
\label{methodology}


\subsection{Theory}
To the best of my knowledge, this is the first theory-based automation exposure index and is based on \citet{erdilMoravecsParadoxIts2024}. I consider four factors that contribute to the general automatizability of a given task, which isn't limited by the current capabilities of AI. These four subhypotheses is Moravec's Paradox proxied by performance variance, data abundance, tacit knowledge, and the algorithmic efficiency gap.

First, Moravec's Paradox suggests that easy tasks are often complex for humans, and vice versa. A potential explanation tested here is that evolutionary optimization can help resolve this paradox. Tasks that are perceived as hard for humans exhibit high variance between them, as they were developed late in evolution, such as mathematics and coding. \citet{mitchell2025visual} deliver neurobiological evidence that late-evolved skills show higher variability in the location of tasks in the brain. Meanwhile, tasks that are easy for humans, such as somatosensory tasks, exhibit low variance for humans but are more challenging for AI, as humans have had more time during evolution to optimize their performance. Thus, performance variance serves as a proxy for the evolutionary optimization explanation of Moravec's Paradox. 

Second, data abundance predicts how well AI can perform a task. If a model is trained on more data of a specific task, it gets better at it. This also holds for general-purpose AI that is trained on a broad dataset and can perform a wide range of tasks. Reasoning models are increasingly trained on high-quality human data, showing the continuing importance of data.

Third, tacit knowledge, sometimes called implicit knowledge, means that the task requires something hard to verbalize, such as intuition, wisdom, and experience.
Thus, it might be hard to teach AIs if people can't communicate with each other. \citet{autorPolanyisParadoxShape2014} calls this Polanyi's Paradox and identifies it as a main driver of labor market outcomes. Furthermore, \citet{bresnahanWhatInnovationPaths2024}also sees tacit knowledge as an obstacle to automation. Workers tend to learn this on the job, and thus this factor varies the most between junior and senior employees. 

Fourth, the human brain is inherently better developed for some tasks than neural networks. While AIs are trained on more compute and data already, they don't yet outperform humans across the board, indicating that the brain retains some algorithmic advantages. I refer to this as the algorithmic efficiency gap. Humans learn better from limited data and are more sample-efficient\citep{Lake2017Building}. For instance, the human brain can better process somatosensory inputs, whereas it does not have any algorithmic advantage in reading and summarizing large amounts of text. In fact, AIs can already perform this task faster with a decent level of quality.

I hypothesize that all four factors can explain automatibility and can serve as an addition to capability-based exposure indices, that might get outdated sooner. I will refer to these factors as subhypotheses at times.

\subsection{Data}

My primary dataset is over 19,000 task statements and over 900 occupations from O*NET that cover the whole US labor market. This dataset from the US Department of Labor is widely regarded as the best source for neutral and comprehensive labor market coverage, and multiple prior indices have used it. The fine-grained tasks of the dataset provide a comprehensive overview of the entire labor market. 

To bring the exposure to the labor market, I harness the Occupational Employment and Wage Statistics (OEWS) from the US Bureau of Labor Statistics. This dataset delivers high-quality and extensive data, and I use the most recent 2024 series and the 2021 series for comparison with other indices, primarily Eloundou et al. (2024).

For the validation, I also use the indices of  \citet{Brynjolfsson2018}), \citet{feltenOccupationalIndustryGeographic2021}, \citet{Webb2019}, \citet{eloundouGPTsAreGPTs2024}, and \citet{FREY2017254}. I follow the standardization method of Eloundou et al. (2024), who also thankfully shared their data with me. 
However, most of the data was created by me through AI annotation, as described in the following sections.


\subsection{Prompting and Weighting}
I granted the AI access in the user prompt to the occupation and the detailed O*NET task statement, which provides a fine-grained description of the tasks the AI performs. The system prompt contains explicit annotation advice, including a detailed description of the criterion, the description of each exposure level, and an example for each level of exposure. This basic structure, which includes a 0-2 ranking, is based on Eloundou et al. (2024). In contrast to GPT-4's time reduction under tool usage, the exposure levels don't have a clear interpretation. But what causes the limited ranking from 0-2 - the lack of scope sensitivity of models - also might make the 50 times reduction difficult for LLMs to comprehend.  Appendix \ref{app} shows the whole system prompt. 

O*NET classifies tasks as core to an occupation when they're relevant and essential to an occupation. Following prior literature, I weight these core tasks twice for occupational indices. 

\subsection{Aggregation}
I average the hypotheses with equal weighting and also average the output of the three AI models, GPT o4-mini, Claude 4 Sonnet, and Gemini 2.5 Flash. This reduces idiosyncratic model bias, making the results more robust. The success rate of Claude and OpenAI is quite high, but for Gemini, it is around 60\%. I restrict the analysis to occupations with successful annotations from at least two models and partly reduce the number of tasks to compare them on the SOC-6 level with other indices, which fuse some occupations (N=681).   I only include functions in the analysis that have at least two models. All three models were reasonably close to the state of the art, but Gemini 2.5 Flash is the fastest and least capable model, standing out as an outlier on benchmarks.

All four subhypotheses could serve as independent indices and are treated as such during the validation.

\begin{figure}[h]
  \centering
  \includegraphics[width=0.8\textwidth]{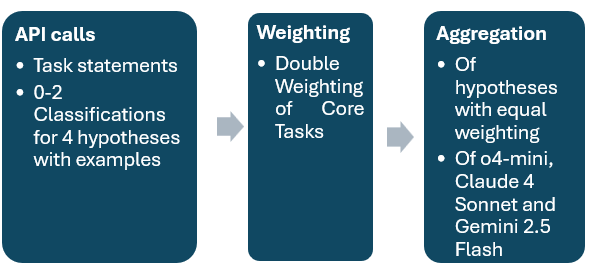}
  \caption{Overview of Steps in Data Creation}
  \label{fig:meth_overview}
\end{figure}

The main specification for the annotation is: 
\begin{equation}
E_j = \frac{\sum_{i} w_i \cdot 0.25 \cdot (PV_i + TK_i + DA_i + AG_i)}{\sum_{i} w_i}
\label{eq:exposure}
\end{equation}
PV, TK, DA, and AG represent the four subhypotheses on the task level i, which I explained in more detail in the theory section. w represents the weight, which is double if it's a core task. I also normalize to the span 0-2, such that the left-hand side, E, represents the average overall automation score at the occupation level j.
The relationship to other exposure indices and to a simple labor market benchmark is consistent with this predictive approach. 



\section{Results}
\label{results}


\subsection{Summary Statistics}

The following occupations have the highest and lowest exposure in the database. 
\begin{table}[h]
\caption{Occupations with the most extreme scores}
\centering
\begin{tabular}{l|l}
\textbf{Highest Exposure} & \textbf{Lowest Exposure} \\
\hline
Online Merchants & Riggers \\
Logistics Analysts & Electrical Power-Line Installers and Repairers \\
Graphic Designers & Rock Splitters, Quarry \\
Market Research Analyst & Refractory Materials Repairers \\
Pharmacy Technicians*& Roustabouts, Oil and Gas \\
\end{tabular}
\end{table}
\footnote{Pharmacy Technicians showed the highest model disagreement and may represent a classification anomaly. }
This suggests that digital tasks are most exposed to pattern recognition. Anthropic has experimented with an autonomous online merchant, which yielded mediocre results with promising signs. Physical tasks show the lowest exposure. Pharmacy Technicians might be the odd ones out here, but they showed the highest model disagreement and may represent a classification anomaly. 
Via text recognition, I found that management, STEM, and sciences are the most exposed job categories, while maintenance, agriculture, and construction are the least exposed categories. I also grouped the over 900 occupations into 10 job categories, and the following table shows the results. 


\begin{table}[h]
\caption{Job categories with the most extreme scores}
\centering
\begin{tabular}{l|l}
\textbf{Highest Exposure} & \textbf{Lowest Exposure} \\
\hline
Management& Maintenance\\
STEM& Agriculture\\
Sciences& Construction\\
\end{tabular}
\end{table}

\subsection{Wages and Employment}
The relationship of exposure with wages is positive. Initially, higher salaries are associated with greater exposure, which is consistent with wage compression resulting from AI and the increased exposure of highly educated jobs. Exposure may lead to the best-paying jobs, potentially due to the development of judgment skills (Gans et al.). Still, not even the means are significantly falling, so this has to be interpreted cautiously.

\begin{figure}[h]
  \centering
  \begin{minipage}{0.48\textwidth}
    \centering
    \includegraphics[width=\textwidth]{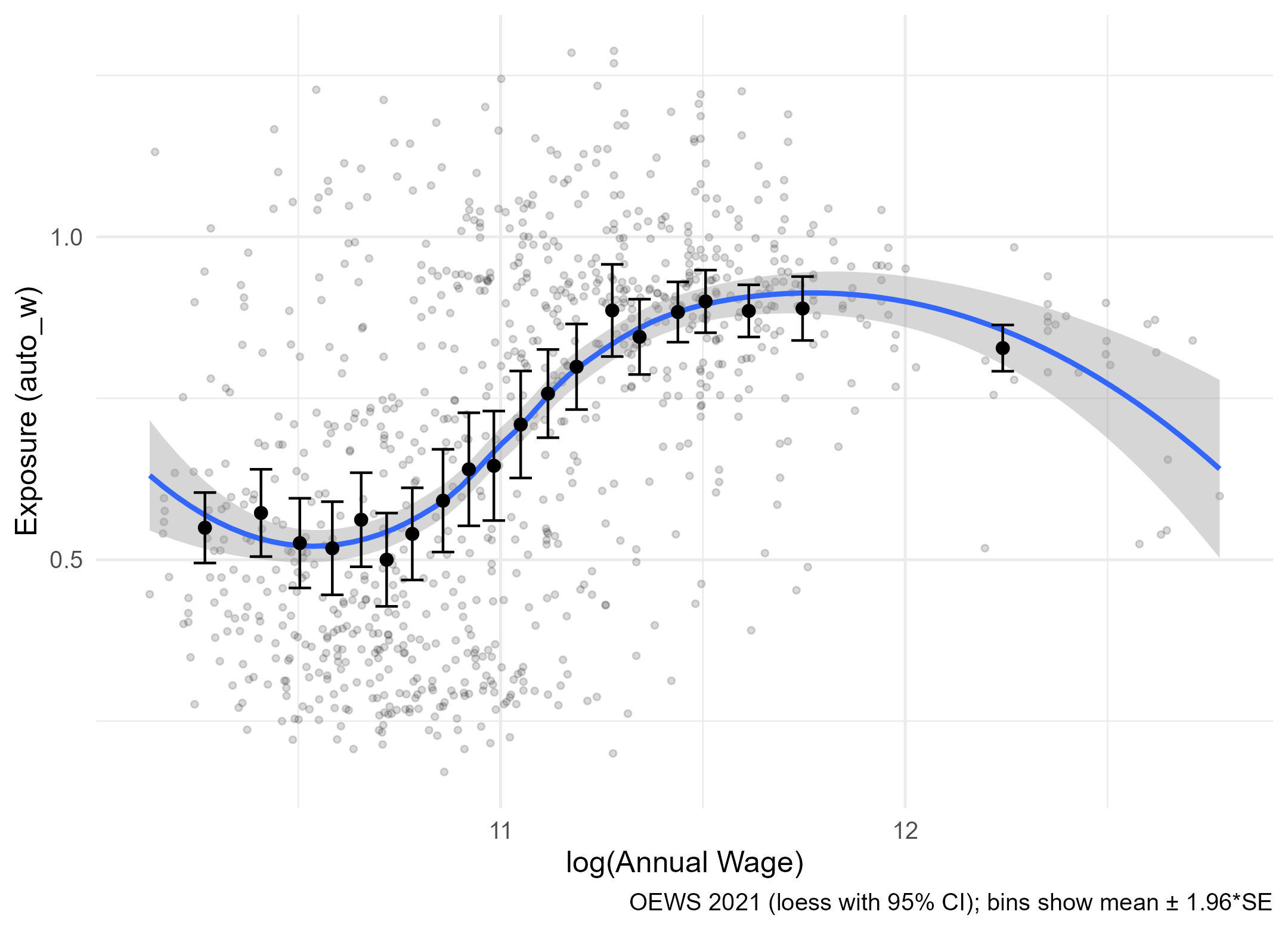}
    \caption*{2021} 
  \end{minipage}\hfill
  \begin{minipage}{0.48\textwidth}
    \centering
    \includegraphics[width=\textwidth]{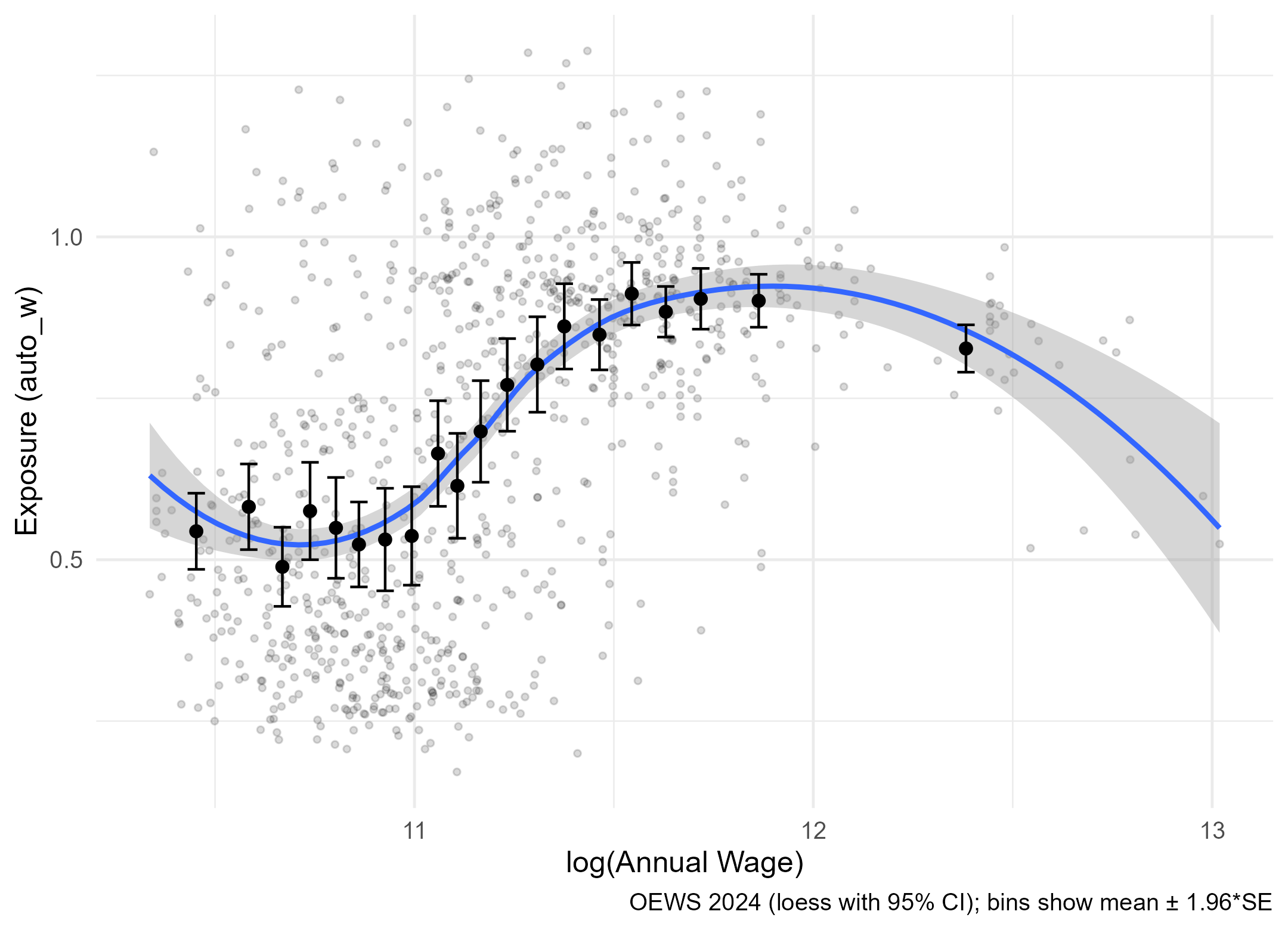}
    \caption*{2024}
  \end{minipage}
  \caption{Largely positive relationship between wages and exposure with confidence interval for the mean}
  \label{fig:wage_curve}
\end{figure}
The binscatter plots depict the exposure to language models (LLMs) in various occupations. Employment and wage data are sourced from the BLS-OEWS survey conducted in May 2021 (for comparison with other indices, such as \cite{eloundouGPTsAreGPTs2023}, which is highly consistent, and May 2024 for more recent data. The appendix \ref{app: Wages and Subhypotheses for 2024} shows for each subhypothesis the corresponding wage plot. 

\begin{figure}[h]
  \centering
  \begin{subfigure}{0.48\textwidth}
    \centering
    \includegraphics[width=\textwidth]{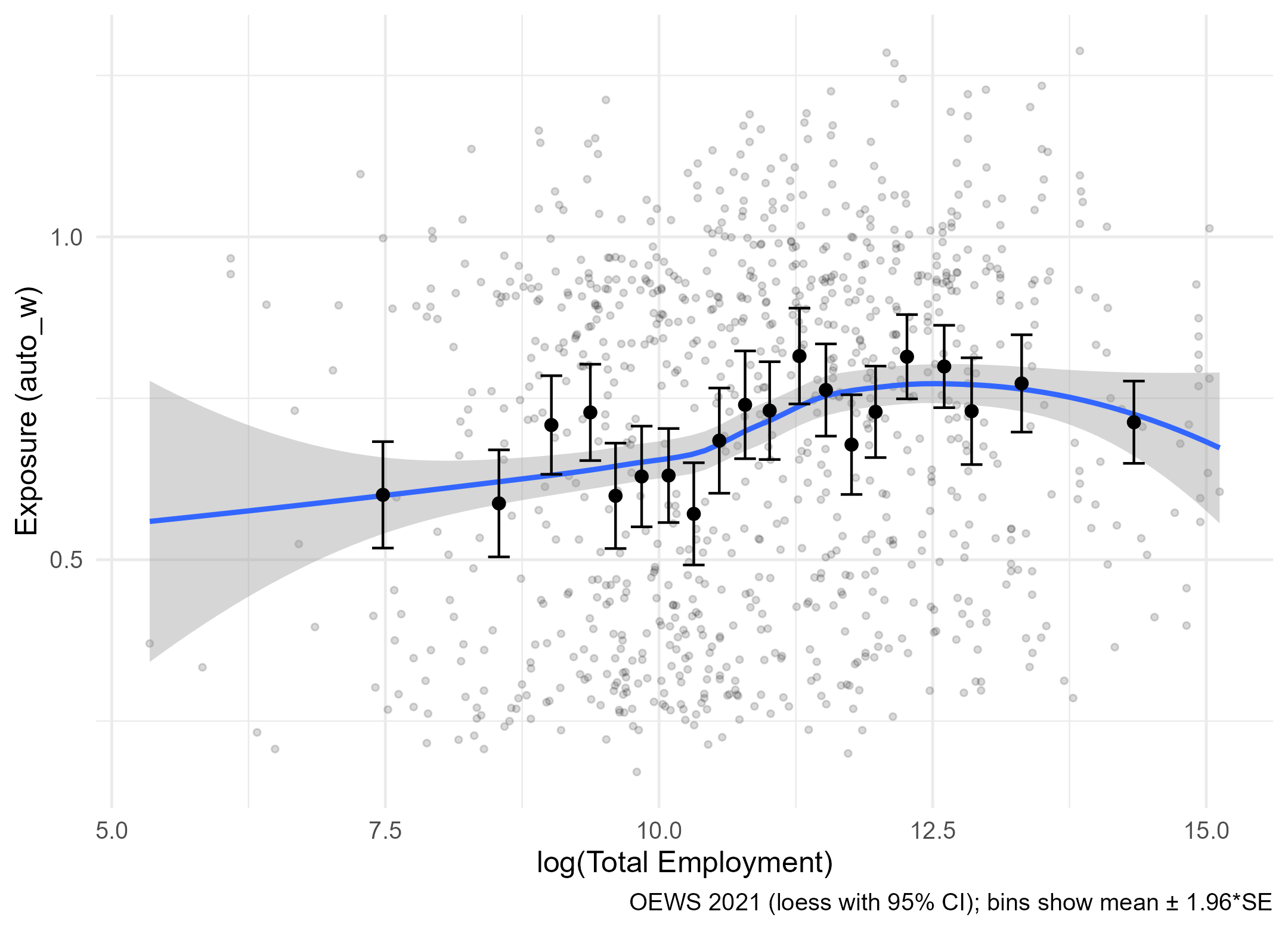}
    \caption{2021}
  \end{subfigure}\hfill
  \begin{subfigure}{0.48\textwidth}
    \centering
    \includegraphics[width=\textwidth]{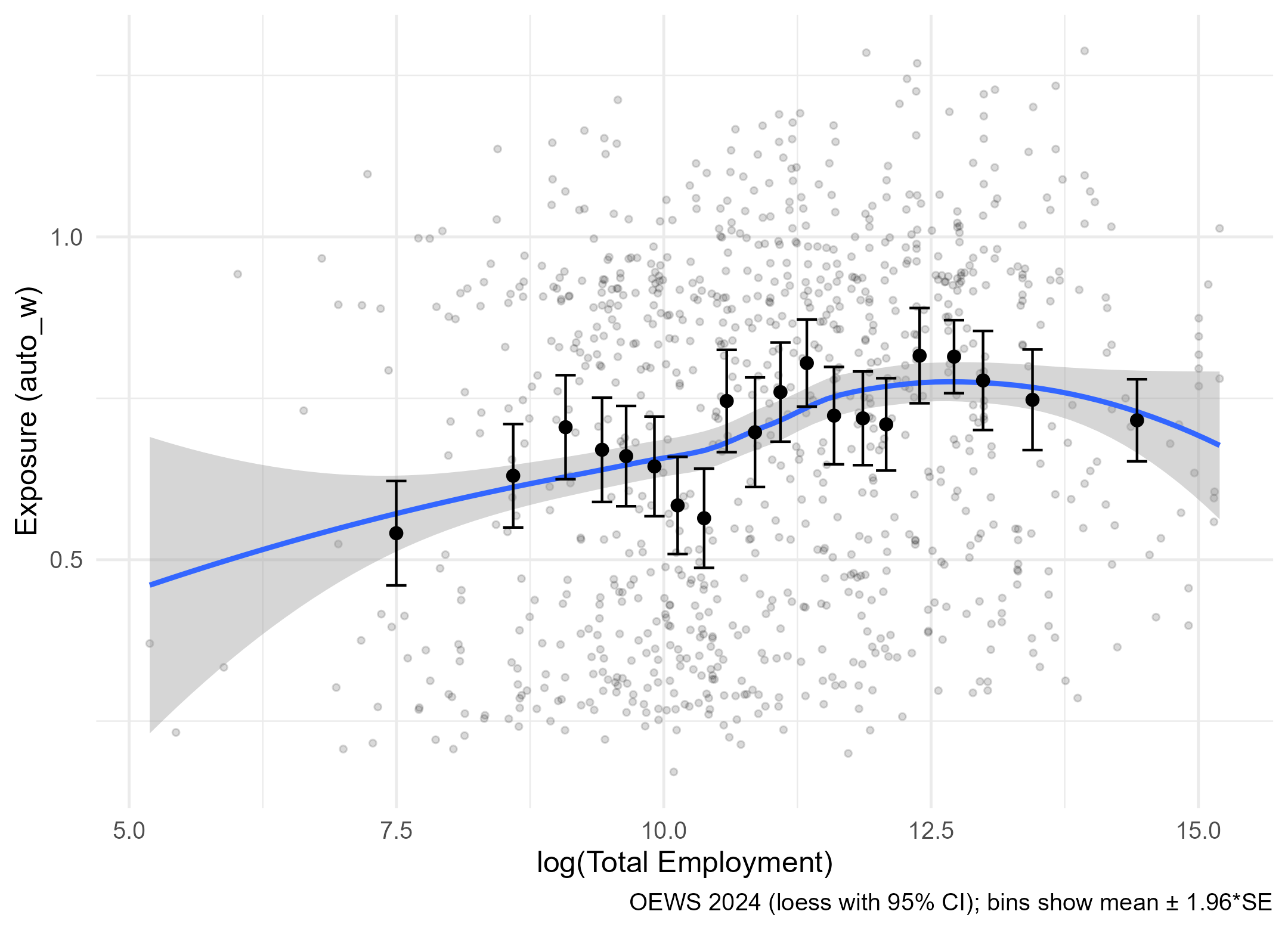}
    \caption{2024}
  \end{subfigure}
  \caption{Relationship between exposure and log employment on occupation level (2021 vs 2024)}
  \label{fig:logemp_exposure}
\end{figure}
As in prior literature, there's no clear relationship between the number of workers employed in an occupation and its AI exposure, with potentially greater exposure in more common occupations. This suggests that the share of workers affected by automation would be at least as high as the share of occupations, but a more detailed interpretation isn't warranted. Appendix \ref{app: Eloundou_labor market} shows that the relationship of my index with labor market data is consistent with Eloundou's et al.(2024) estimates for both employment and wages. This holds for both their AI-annotated and human-rated data. This remarkable, typical pattern also shows up in the direct comparison with other indices in the following section. This validates both the AI annotation approach due to the similarity with human ratings, and cross-validates their two approaches. As will be seen in the next chapter, this isn't the only point of consistency and cross-validation between Eloundou et al. (2024)'s measure and mine. 









\section{Validation of Index}
\label{Validation of Index}
\subsection{Comparison to Earlier Efforts}
This paper aims to contribute to the diverse literature on AI exposure indices. Previous studies have employed various methods and yielded diverse results, as outlined in earlier sections. The correlation with Eloundou et al. (2024) is particularly high, with 0.72, as the basic methodology is similar. Nonetheless, the completely different prompts exhibit such a high correlation, which validates the standard AI annotation methodology, as the exposure indices seem robust to prompting. The regression table indices indicate that each subhypothesis, on its own, is a valid exposure index, given the substantial share of variance it can explain. It suggests that automation exposure is very multi-dimensional, as even the overall index provides little additional explanation. This is consistent with a considerable difficulty in finding proper unidimensional indices. The overall index is negatively correlated with Webb's software and robot indices, while being positively correlated with the AI index. It is negatively correlated to the three pre-LLM AI exposure indices of Frey and Osborne (2017), Felten et al. (2021), and Brynjolfsson et al. (2017). It's negatively correlated to routine manual jobs, but positively associated with routine cognitive, which is consistent with the summary statistics in the previous sections, and the job categories shown there. The positive relationship to Eloundou et al's preferred beta measure is encouraging. 


\begin{table}[!htbp] \centering 
  \caption{Regression of Automation Measures on Prior Literature} 
  \label{} 
\resizebox{\textwidth}{!}{%
\begin{tabular}{@{\extracolsep{5pt}}lccccc} 
\\[-1.8ex]\hline 
\hline \\[-1.8ex] 
 & \multicolumn{5}{c}{\textit{Dependent variable:}} \\ 
\cline{2-6} 
\\[-1.8ex] & Overall Index & Productivity Variance & Data Abundance & Tacit Knowledge & Algorithmic Gap\\ 
\\[-1.8ex] & (1) & (2) & (3) & (4) & (5)\\ 
\hline \\[-1.8ex] 
 Software (Webb) & $-$0.00012 & $-$0.00178$^{***}$ & 0.00057 & 0.00103$^{*}$ & 0.00118$^{**}$ \\ 
  & (0.00040) & (0.00045) & (0.00045) & (0.00043) & (0.00043) \\ 
  & & & & & \\ 
 Robot (Webb) & $-$0.00169$^{***}$ & $-$0.00156$^{**}$ & $-$0.00182$^{***}$ & 0.00149$^{**}$ & $-$0.00212$^{***}$ \\ 
  & (0.00044) & (0.00049) & (0.00049) & (0.00048) & (0.00047) \\ 
  & & & & & \\ 
 AI (Webb) & 0.00122$^{***}$ & 0.00217$^{***}$ & 0.00086$^{*}$ & $-$0.00157$^{***}$ & 0.00153$^{***}$ \\ 
  & (0.00035) & (0.00040) & (0.00040) & (0.00038) & (0.00038) \\ 
  & & & & & \\ 
 Suitability for ML & $-$0.20122$^{***}$ & $-$0.66181$^{***}$ & 0.51138$^{***}$ & 0.90404$^{***}$ & 0.71782$^{***}$ \\ 
  & (0.05236) & (0.05885) & (0.05877) & (0.05664) & (0.05641) \\ 
  & & & & & \\ 
 Routine Cognitive & 0.04717$^{***}$ & $-$0.07199$^{***}$ & 0.05422$^{***}$ & 0.03645$^{***}$ & 0.05279$^{***}$ \\ 
  & (0.00859) & (0.00965) & (0.00964) & (0.00929) & (0.00925) \\ 
  & & & & & \\ 
 Routine Manual & $-$0.04415$^{***}$ & $-$0.01220 & $-$0.06786$^{***}$ & 0.01039 & $-$0.00869 \\ 
  & (0.01173) & (0.01318) & (0.01316) & (0.01269) & (0.01264) \\ 
  & & & & & \\ 
 AI Exposure (Felten) & $-$0.03137$^{*}$ & 0.00260 & 0.04476$^{**}$ & $-$0.11142$^{***}$ & 0.00775 \\ 
  & (0.01300) & (0.01461) & (0.01459) & (0.01406) & (0.01401) \\ 
  & & & & & \\ 
 Frey-Osborne Automation & $-$0.16016$^{***}$ & $-$0.16729$^{***}$ & $-$0.07953$^{**}$ & 0.10097$^{***}$ & 0.05432 \\ 
  & (0.02622) & (0.02947) & (0.02943) & (0.02836) & (0.02825) \\ 
  & & & & & \\ 
 Eloundou $\beta$ & 0.50409$^{***}$ & 0.17899$^{**}$ & 0.74291$^{***}$ & 0.18811$^{***}$ & 1.06082$^{***}$ \\ 
  & (0.05121) & (0.05756) & (0.05747) & (0.05539) & (0.05517) \\ 
  & & & & & \\ 
 Constant & 1.32248$^{***}$ & 2.94626$^{***}$ & $-$0.86286$^{***}$ & $-$1.79062$^{***}$ & $-$1.93400$^{***}$ \\ 
  & (0.17469) & (0.19635) & (0.19607) & (0.18896) & (0.18822) \\ 
  & & & & & \\ 
\hline \\[-1.8ex] 
Observations & 681 & 681 & 681 & 681 & 681 \\ 
R$^{2}$ & 0.65489 & 0.71368 & 0.67720 & 0.74663 & 0.75309 \\ 
Adjusted R$^{2}$ & 0.65026 & 0.70984 & 0.67287 & 0.74323 & 0.74978 \\ 
Residual Std. Error (df = 671) & 0.16240 & 0.18254 & 0.18228 & 0.17568 & 0.17498 \\ 
F Statistic (df = 9; 671) & 141.47940$^{***}$ & 185.83410$^{***}$ & 156.40830$^{***}$ & 219.69980$^{***}$ & 227.39810$^{***}$ \\ 
\hline 
\hline \\[-1.8ex] 
\textit{Note:}  & \multicolumn{5}{r}{$^{*}$p$<$0.05; $^{**}$p$<$0.01; $^{***}$p$<$0.001} \\ 
 & \multicolumn{5}{r}{Standard errors at 6digit SOC level  in parentheses.} \\ 
 & \multicolumn{5}{r}{Webb measures on 0-100 scale. Routine measures standardized.} \\ 
\end{tabular} 
}
\end{table} 
This table shows the regression of my exposure scores on prior measures of occupational exposure to AI and automation. Each measure is kept in its original scale, with the exception of routine cognitive and routine manual scores from  (Acemoglu and Autor, 2011). Those two scores are standardized to have a mean of zero and a variance of 1. I also look at the correlation with the various indices separately in the following correlation triangle: 

\begin{figure}[h]
\centering
\includegraphics[width=1\textwidth]{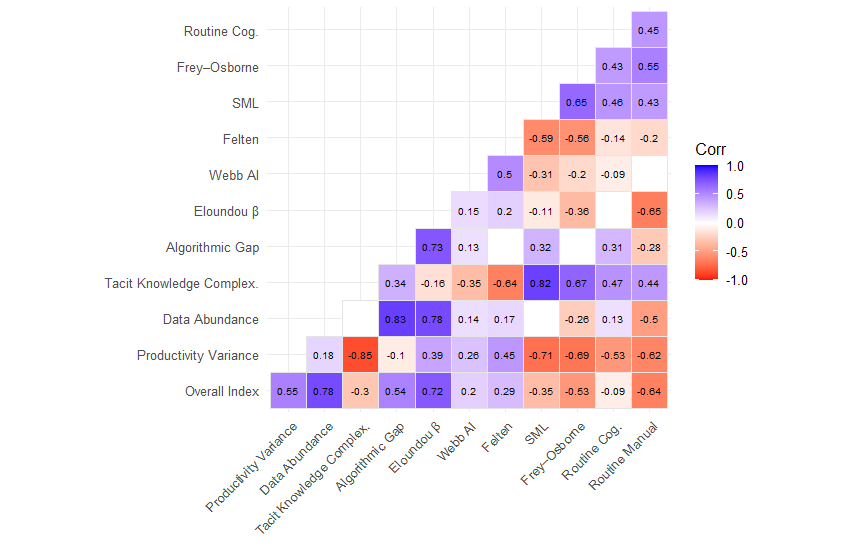}
\caption{Correlations with various exposure indices }
\label{fig:correlations}
\end{figure}


Naturally, most subhypotheses contribute positively to the overall index as can be seen in the lowest row. Tacit knowledge is negatively correlated with the other indices and thsu contributes negatively to the overall index (has a negative beta). 
The correlation with Eloundou et al. (2024) is particularly high, with 0.72, as the basic methodology is similar. Nonetheless, the completely different prompts exhibiut such a high correlation, which validates the standard AI annotation methodology as the exposure indices seem robust to prompting. 
The indices are largely ranked temporarily, and it's easy to visually see that more recent efforts correlate strongly with previous efforts as the triangle transitions from blue to red when the distance from the hypotenuse increases. This also confirms that indices based on current capabilities are fast outdated; only time will tell whether this theory-based index will be helpful for longer.

\subsection{Comparison between models}
I also use multiple AI models for annotation, which have slightly different opinions on automatibility. The reasoning models of OpenAI and Anthropic have a very high correlation of 0.81, but in general, the correlations are pretty high, validating the AI annotation approach. Flash is a smaller and faster model and serves as the outlier, with Claude scoring jobs often in between OpenAI and Gemini. The correlation with itself is also below ,one, as it depends on the random seed when the temperature is above 0.

\begin{table}[h]
\centering
\caption{Correlation triangle between models}
\label{tab:correlation_models}
\begin{tabular}{lccc}
\hline
\textbf{Correlation} & \textbf{OpenAI GPT o4-mini} & \textbf{Gemini 2.5 Flash} & \textbf{Claude 4 Sonnet} \\
\hline
OpenAI GPT o4-mini   &    &      &      \\
Gemini 2.5 Flash     & 0.60 &    &      \\
Claude 4 Sonnet      & 0.81 & 0.71 &     \\
\hline
\end{tabular}
\end{table}

Here, pharmacy technicians are the job with the highest model disagreement, and in general, it's the smallest model flash that produces the 15 most prominent outliers. This is consistent with scaling laws also holding for this classification task.  The specific examples provided may help clarify this. 
\begin{figure}[h]
  \centering
  \includegraphics[width=1\textwidth]{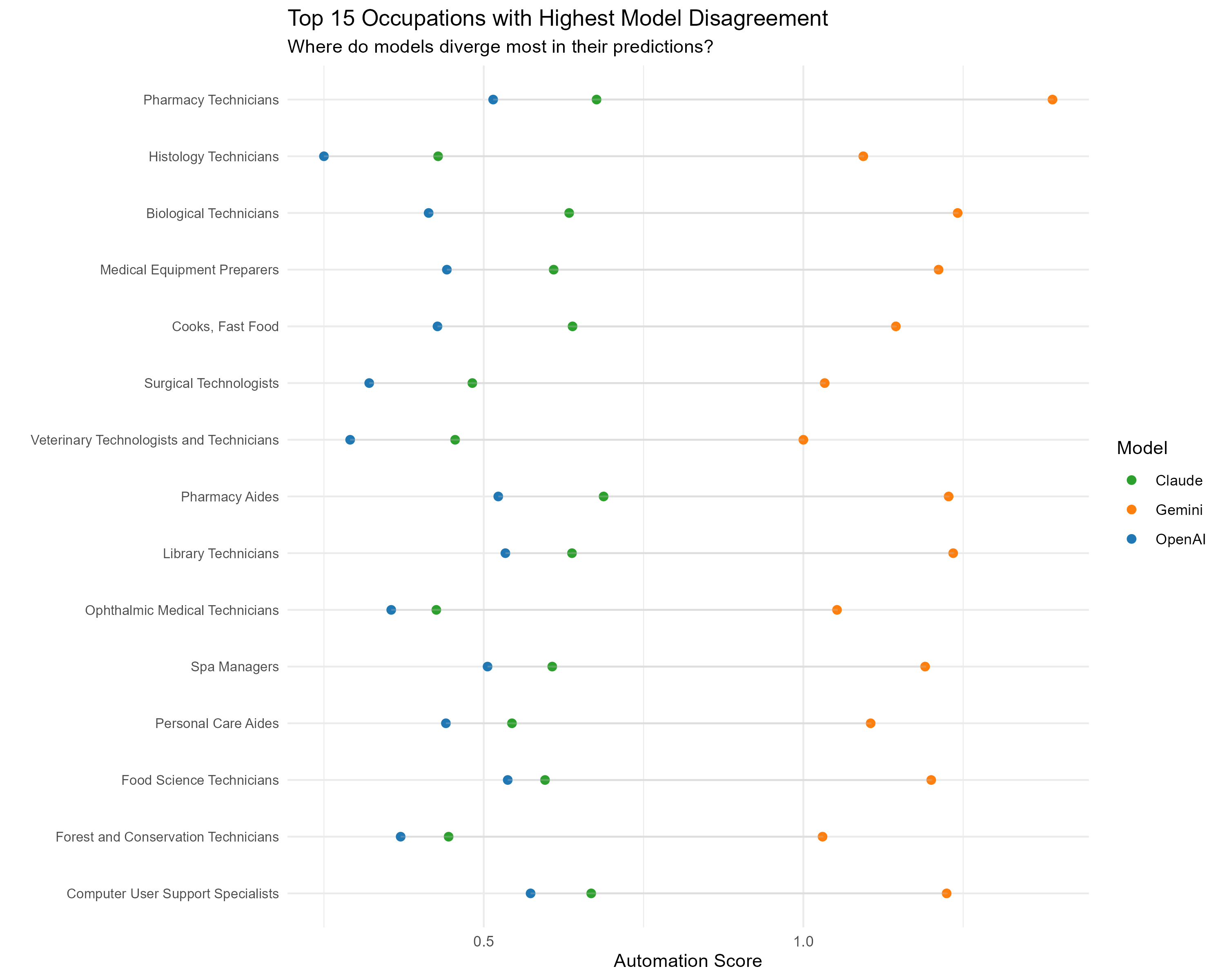}
  \caption{Top disagreement across occupations}
  \label{fig:top_disagreement_occupations}
\end{figure}

\begin{figure}[h]
  \centering
  \includegraphics[width=1\textwidth]{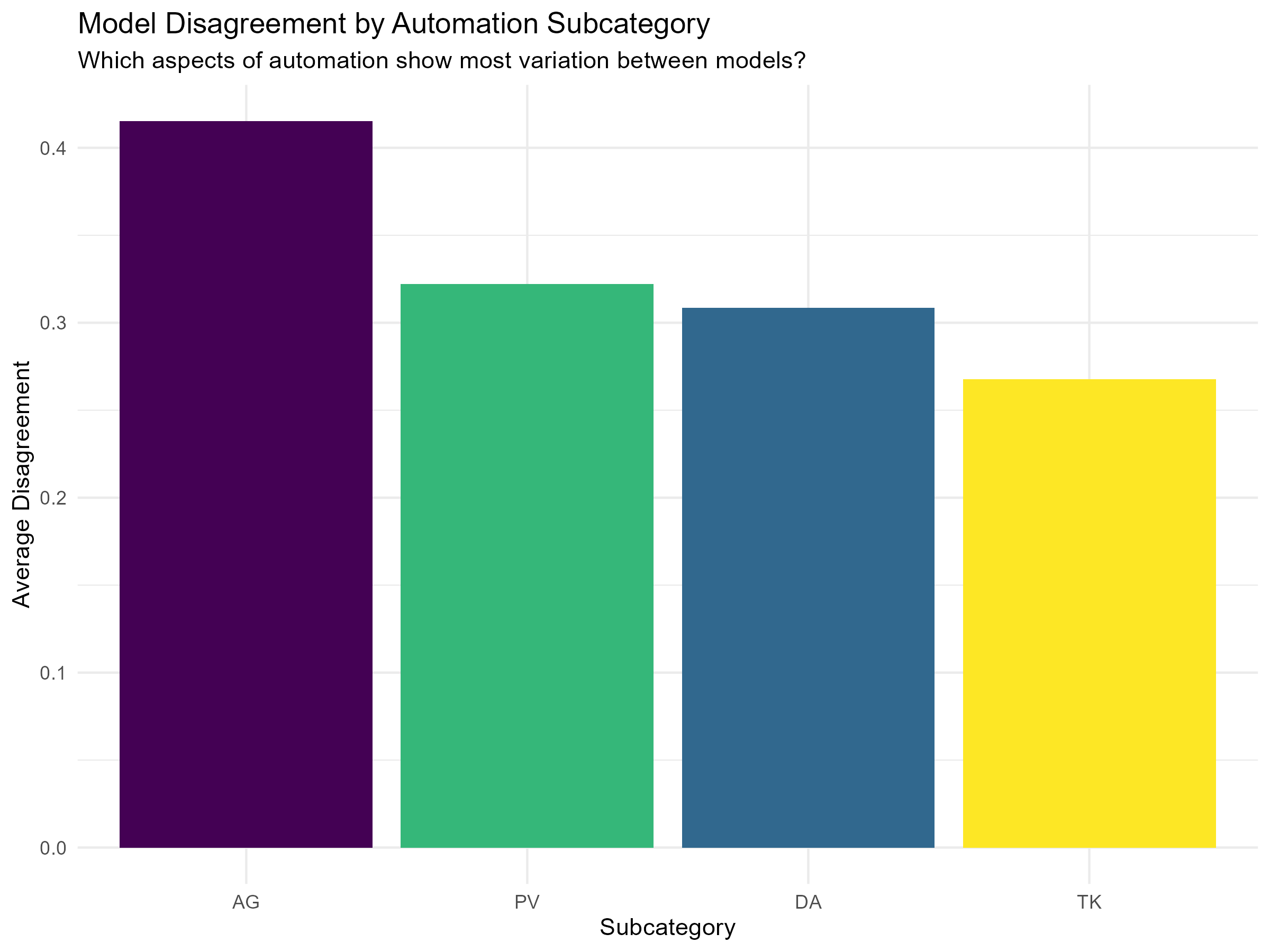}
  \caption{Disagreement across subcategories}
  \label{fig:subcategory_disagreement}
\end{figure}
The average disagreement between models is the largest for the algorithmic efficiency gap, but not by much; still, a better definition might have been most helpful here. The anticorrelated tacit knowledge shows the lowest disagreement, which makes this outlier result more robust. In general, Gemini seems to classify technicians in unusual environments such as pharmacy, library, and ophthalmology. This indicates that reasoning models can abstract better away from naming conventions and take the context better into account.

\subsection{ Validation of Theory}
This index closely follows the theoretical predictions by \citet{erdilMoravecsParadoxIts2024}. Thus, it's a faithful implementation of its theory and the four subhypotheses. This piece predicts four broad tasks that will be automated earlier: scientific research, software engineering, art without sensomotorical skills, and management. This is highly consistent with the three job categories with the highest exposure in Table 2 (management, STEM, and science). Furthermore, a graphic designer is the third most exposed occupation and probably archetypal for "art without sensomotorical skills". This might partly be caused by anchoring bias from the system prompt, for example, as debugging software is given as an example for high data abundance, and writing technical documentation for software is an algorithmic advantage that humans have. This indicates that superficial similarities to the prompt, as seen in \ref{app:prompt}, might contribute to the high exposure classification of software engineering. However, these don't even feature among the most exposed occupations (although Logistics and Market Research could be seen as related). Software engineering falls under STEM jobs, but overall, the high consistency beyond software engineering is hardly driven by these classification examples superficially. Tacit knowledge, which is the primary explanation for graduate unemployment, exhibits the highest covariance with previous indices. However, tacit knowledge also shows a highly positive relationship with wages, which is consistent with the notion that highly educated jobs are among the first to be automated.


\section{Discussion}
\label{Discussion}
\subsection{Validity of Methodology}





The general methodology of AI annotation pioneered by Eloundou et al. (2024) holds up quite well and seems robust to the concrete specification, as various results demonstrate. The theory-based approach might be a good combination of the strengths of expert surveys and a systematic O*NET-based approach, but the results are too preliminary for a final judgment. The methodology is also robust to the choice of models, though reasoning models might perform slightly better. 

These results are consistent with Moravec's Paradox and its explanation as evolutionary optimization. It shows the potential for theory-based exposure indices that might be more robust and useful for a longer time. If AI were substitutive, it would indicate a wage compression and a fall in inequality, holding capital income constant. This has interesting political economy implications: Richer, higher-educated voters might have more to lose from AI adoption, which might exacerbate polarization on the topic of AI.  

The validity of the classification itself is also controversial. The LLMs might also classify based on rudimentary commonalities between the job name and the subcategory at hand. Indeed, Gemini 2.5 Flash shows this in an egregious way and classifies all jobs with the name technicians, whether pharmacy, biological, histology, medical, food science, library or forest technicians, as much more exposed than the reasoning models. In general, classification is a much easier and much better-defined task than those in the O*NET database, and the AI classification seems somewhat useful and follows the classification examples. It's also consistent with rudimentary human classification, but further research should investigate this. Eloundou et al. (2024) showed the consistency of AI-nad human-based annotation. 

Future research could examine the expertise required in this job and thus attempt to create a more detailed picture of the ambivalent effects automation has on jobs, which differ depending on whether low-skilled tasks or high-skilled tasks of an occupation are automated \citep{AutorThompson2025Expertise}. This would require no tonly the calculation of average exposure of an occupation but also the ranking of tasks regarding the required expertise.

Researchers could use these to study labor market impact across different time periods, as the index isn't based on specific model capabilities. All the data are open-sourced and can be used for this. Studies deploying other indices could be easily replicated as long as they use the O*NET classification

Future research could also address some limitations of this study. When examining labor market outcomes with exposure, it remains a challenge to avoid endogeneity concerns. However, some studies discussed in the following subchapter attempt to overcome this by controlling for firm-level shocks, using difference-in-difference and triple-difference estimations.  At times, the theoretical grounding of the hypotheses hasn't been validated rigorously before. Additionally, further research would be needed to investigate the causality of exposure on labor market outcomes. The treatment of Gemini's outcome, which changes some estimations, such as for Pharmacy Technicians, significantly, could introduce sample selection bias, as Gemini only classified 60\% tasks. A direct verification with human annotators would strengthen the credibility of the AI annotation; however, Eloundou et al. (2024) demonstrated the validity of the methodology. The weighting of the four factors could have also followed another pattern, but the separate treatment of the factors allows for the exploration of the differences. 
Overall, creating a meaningful AI automation exposure index remains a challenging task. Still, AI annotation doesn't seem to perform worse than other methods, especially older pre-LLM ones. Nonetheless, it remains to be seen how temporally stable this exposure index is.

\subsection{Tacit Knowledge and Early-Career Unemployment}
Recent research validates the usefulness of the post-LLM AI annotation method as a tool for predicting the impact of AI on the labor market. This is to my knowledge the first evidence for the adverse effect of AI exposure on the labor market. Both \citet{lichtingerGenerativeAISeniorityBiased2025} and \citet{brynjolfssonCanariesCoalMine}  
hypothesize that tacit knowledge, which is learned on the job and hard to verbalize, explains the seniority-biased technological change. In my framework, tacit knowledge is the only factor that is negatively correlated with wages as Appendix \ref{app: Wages and Subhypotheses for 2024} shows. This is consistent with two strands of literature: A general higher exposure of highly-educated workers to AI compared to lower-educated ones, as low-skilled workers see higher productivity boosts from AI \citep{Noy2023}, indicating complementarity. On the other hand, highly educated but not yet very highly skilled job entrants with a college degree have less tacit knowledge that would protect them from the substitutive effect of AI. This is consistent with my finding that higher-paid workers have more tacit knowledge. This factor is also the only one that shows a negative relationship with wages and is negatively correlated with the other subhypotheses. This could mean that seniority-biased technological change could be followed by reverse skill-biased technological change over time, if AI exposure in other domains also becomes substitutive. 
If the adverse effects on early-career workers persist, this would also have long-term negative effects on the intergenerational transmission of knowledge. \cite{ideAutomationAIIntergenerational2025} demonstrates that if the lack of on the job learning opportunities continues, there might be less labor supply for senior roles in the future, which could reduce productivity growth by 0.05 to 0.35 percentage points.










\section{Conclusion}
\label{conclusion}

This theory-based approach to automation exposure offers multiple key contributions. First, it provides a principled framework grounded in evolutionary theory rather than current technological capabilities. Second, the positive wage relationship suggests AI may reduce inequality by automating high-skill cognitive work while preserving manual jobs, holding capital income constant. Third, the high correlation with Eloundou (0.72) validates the approach, while negative correlations with pre-LLM indices highlight the paradigm shift in automation patterns. Future work should test these predictions as AI capabilities evolve. The high correlations between the models, and especially the reasoning models, are promising, as is the robustness of the AI annotation method to the direct specification. Policymakers can place  at least comparable levels of trust in the AI annotation method as in previous, pre-LLM methods. 

\newpage


\newpage

\appendix

\section*{Appendix}
\label{app}
\appendix

\subsection*{System Prompt}
\label{app:prompt}
"Score each task 0, 1, or 2 (where 2 = most automatable).

Consider the occupation context when scoring. Output JSON: \{"PV": X, "DA": Y, "TK": Z, "AG": W\}

You are an expert economist analyzing tasks for automation potential across four theoretical frameworks. Your analysis will inform academic research on labor market impacts. 

For each task, provide precise scores (0, 1 or 2) in JSON format (where 2= most automatable): \{"PV": X, "DA": Y, "TK": Z, "AG": W\}

\# Performance Variance Taxonomy

Consider how much performance varies between trained professionals when completing this task. Performance variance reflects whether humans have evolved to optimize for this task - high variance suggests the task hasn't been optimized through evolution and may be easier to systematize.

Assume you are evaluating the productivity gap between workers with the same training and experience within the same profession. Consider measurable outputs like speed, accuracy, quality, or efficiency.

Please label the given task according to the taxonomy below.

\#\#
PV0 -- Low variance

Label tasks 0 if professionals show a low productivity gap between workers. These are tasks where most trained professionals perform at similar levels.

\#\# PV1 -- Moderate variance  

Label tasks 1 if there is a medium productivity gap between workers. These tasks show some performance differences but professionals generally achieve comparable results.

\#\# PV2 -- High variance

Label tasks 2 if there is a high productivity gap between workers. These tasks show significant performance differences between professionals, suggesting humans haven't uniformly mastered them.

\#\# Classification examples:

Task: Perform routine blood draws from patients

Label: 0

Explanation: Trained phlebotomists show minimal variance in successful blood draw rates and patient comfort levels.

Task: Evaluate and select providers of services according to customer requirements.

Label: 1

Explanation: Sales professionals show moderate variance in customer satisfaction and sales success rates: some variance in understanding customers

Task: Develop algorithmic trading strategies for financial markets 

Label: 2

Explanation: Quantitative traders show huge variance in returns, with top performers generating consistent alpha while average traders barely beat market indices or lose money

\# Data Abundance Taxonomy

Consider the availability of digital training data for this task. Tasks with massive online datasets of examples can be more easily learned by computational systems.

Think about whether examples of this task being performed exist in digital form online, in databases, or in digital archives. Consider text descriptions, code, documents, or other digital records.

Please label the given task according to the taxonomy below.

\#\# DA0 -- Limited digital data

Label tasks 0 if hardly any digital examples of the task exist online. These tasks are primarily demonstrated physically or have limited digital documentation.

\#\# DA1 -- Moderate digital data

Label tasks 1 if a moderate amount of examples exist digitally. Some documentation and examples are available but coverage is incomplete.

\#\# DA2 -- Abundant digital data  

Label tasks 2 if many examples exist online. These tasks have extensive digital documentation, tutorials, examples, and records.

\#\# Classification examples:

Task: Debug software code and fix errors

Label: 2

Explanation: Billions of code debugging examples exist on GitHub, Stack Overflow, and programming forums.

Task: Collaborate with system architects, software architects, design analysts, and others to understand business or industry requirements.

Label: 1

Explanation: Some digital documentation exists in the form of business requirement documents. some data on collaboration, but enough data on each role

Task: Adjust industrial machinery by feel and sound

Label: 0  

Explanation: This tactile task has minimal digital documentation beyond basic manuals.

\# Tacit Knowledge Taxonomy

Consider how much tacit knowledge, gained through education and especially experience, is required for this task. Tacit knowledge includes judgment, intuition, and skills that are difficult to articulate or codify and those tasks are hard for entry-level graduates to fullfill. 

Evaluate whether the task is typically performed by entry-level workers or requires years of training and seniority.

Please label the given task according to the taxonomy below.

\#\# TK0 -- High tacit knowledge required

Label tasks 0 if they require extensive training, education, and seniority. These tasks are predominantly performed by senior professionals with years of experience.

\#\# TK1 -- Moderate tacit knowledge required

Label tasks 1 if they require some training and education but can be performed adequately by mid-level professionals.

\#\# TK2 -- Minimal tacit knowledge required

Label tasks 2 if they require little training or education and can be performed by entry-level workers following clear procedures.

\#\# Classification examples:

Task: Enter data from paper forms into spreadsheets

Label: 2

Explanation: Requires minimal training beyond basic computer literacy and can be performed by entry-level workers.

Task: Coordinate services for events, such as accommodation and transportation for participants, facilities, catering, signage, displays, special needs requirements, printing and event security.

Label: 1

Explanation: some required to ease coordination, but largely procedural, not judgemental. Coordination is largely procedural - following checklists, timelines, and standard processes. While it requires organizational skills and attention to detail, these can be learned relatively quickly. Mid-level event coordinators routinely handle these logistics. The main skill is project management, not deep expertise or intuition.

Task: Inspect event facilities to ensure that they conform to customer requirements.

Label: 0

Explanation: Need to know where to look to find issues, and how to assess them. Requires extensive experience in event management and facility inspection. Customer requirements are often underspecified and need to be inferred from context. Facilities try to hide issues, so need to be able to spot them.

\# Algorithmic Efficiency Gap Taxonomy

Consider whether this task requires a lot of  physical manipulation, multimodal sensory input, or embodied interaction with the physical world. These requirements create efficiency gaps where humans maintain advantages.

Evaluate if the task could be performed purely through digital interfaces or requires a lot of physical presence and sensory feedback. The human (brain) is more efficient in some domains such as multimodal sensory input, Physical world modelling and embodiment and physical sensoring.

This taxonomy classifies tasks based on domains where the human brain maintains computational efficiency advantages. These advantages stem from our evolved neural architecture's specialized capabilities, not just physical embodiment.

Key Domains of Human Brain Efficiency are Multimodal Integration \& Embodied Cognition, Social-Emotional Processing, Creative Synthesis \& Abstraction, Contextual Flexibility \& Common Sense and Efficient Learning from Sparse Data.

Please label the given task according to the taxonomy below.

\#\# AG0 -- High human brain efficiency advantage

Label tasks 0 if they require significant physical manipulation, multimodal sensory integration (touch, proprioception, spatial awareness), embodied presence, complex social-emotional processing, creative synthesis, or contextual common sense reasoning.

\#\# AG1 -- Moderate human brain efficiency advantage

Label tasks 1 if they involve some physical components, occasional multimodal inputs or some social/contextual interpretation, o

\#\# AG2 -- Minimal human brain efficiency advantage

Label tasks 2 if they can be performed largely through digital interfaces without physical manipulation, multimodal sensing, or significant need for social understanding, creativity, or contextual reasoning.

\#\# Classification examples:

Task: Write technical documentation for software

Label: 2

Explanation: Purely digital task requiring no physical manipulation or sensory input beyond reading and typing.

Task: Direct and coordinate activities of businesses or departments concerned with the production, pricing, sales, or distribution of products.

Label: 1

Explanation: It requires some social skills, but not the main part of the task.

Task: Repair plumbing in residential buildings

Label: 0

Explanation: Requires physical manipulation, spatial reasoning, tactile feedback, and adaptation to unique physical environments.

CRITICAL: Base rankings  on the detailed criteria above. Consider the COMPLETE task description.

IMPORTANT: Base your scores on the task description alone. Consider what the task fundamentally requires, not current technological limitations. Also, do not overanchor on the examples provided, but also give scores between them."

\subsection*{Relationship between Human and AI-annotated Automation Score and Labor Market Data such as Wages and Employment \citep{eloundouGPTsAreGPTs2024}}
\label{app: Eloundou_labor market}

\begin{figure}[h]
  \centering
  \includegraphics[width=0.7\textwidth]{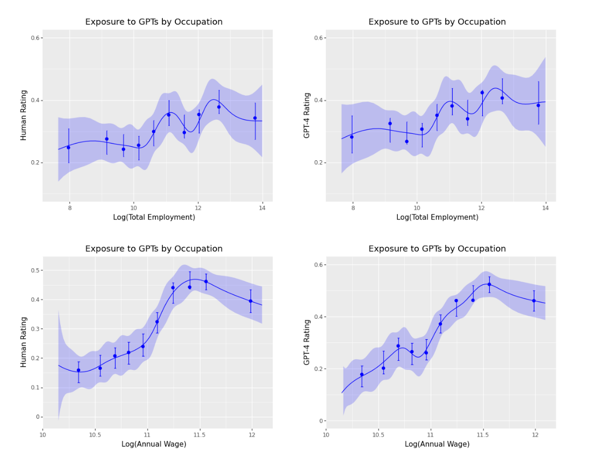}
  \caption{ Positive Relationship between Wages and exposure with insignificant inverted U-curve and no relationship between employment numbers and exposure (Eloundou et al., 2024)}
  \label{fig:eloundou_labor market}
\end{figure}
\newpage

 \subsection*{Wages and Subhypotheses}
\label{app: Wages and Subhypotheses for 2024}

\begin{figure}[h]
  \centering
  \includegraphics[width=0.7\textwidth]{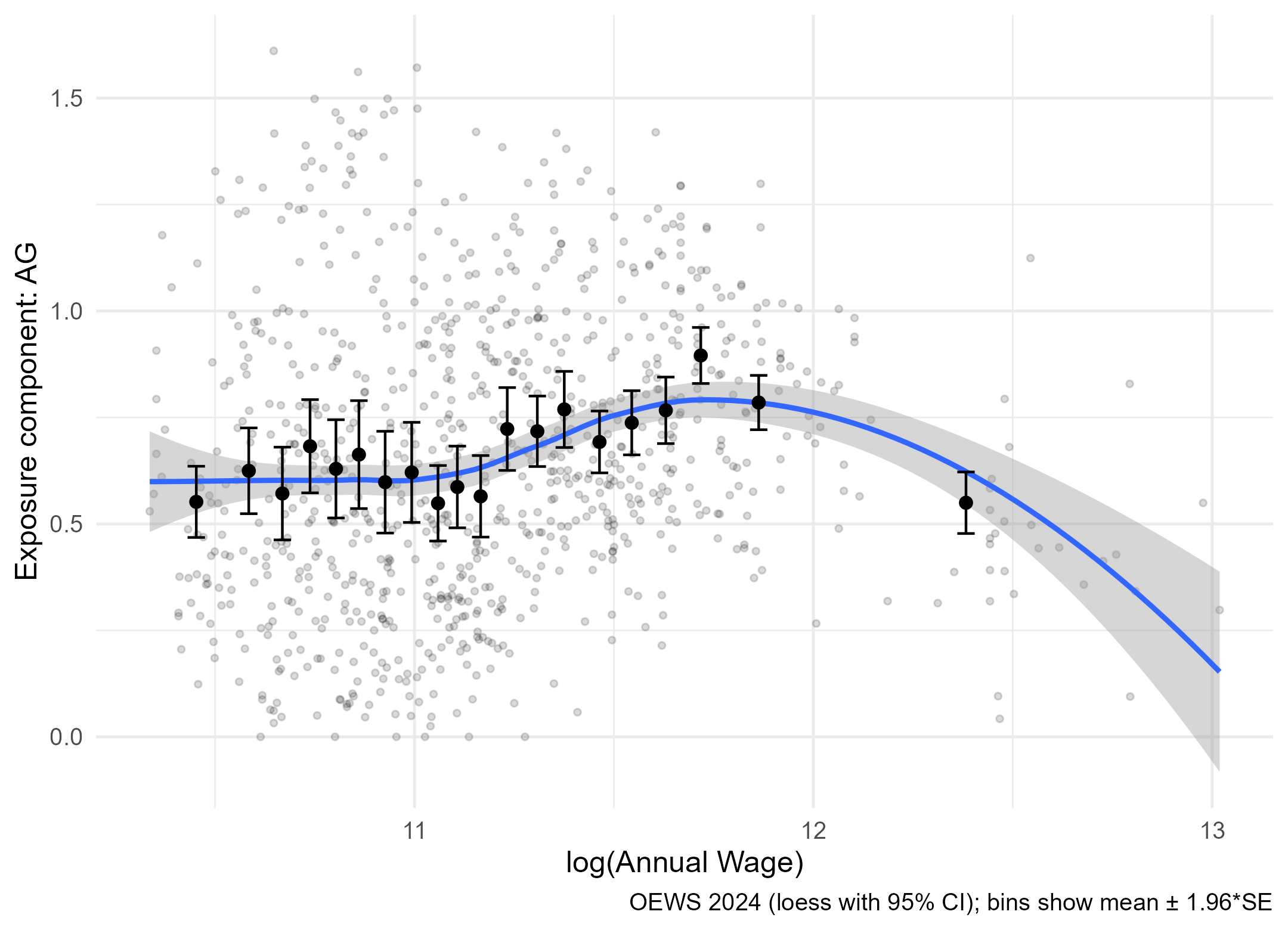}
  \caption{ Algorithmic Efficiency gap (AG) exposure vs log wages, 2024}
  \label{fig:subhyp1}
\end{figure}

\begin{figure}[h]
  \centering
  \includegraphics[width=0.7\textwidth]{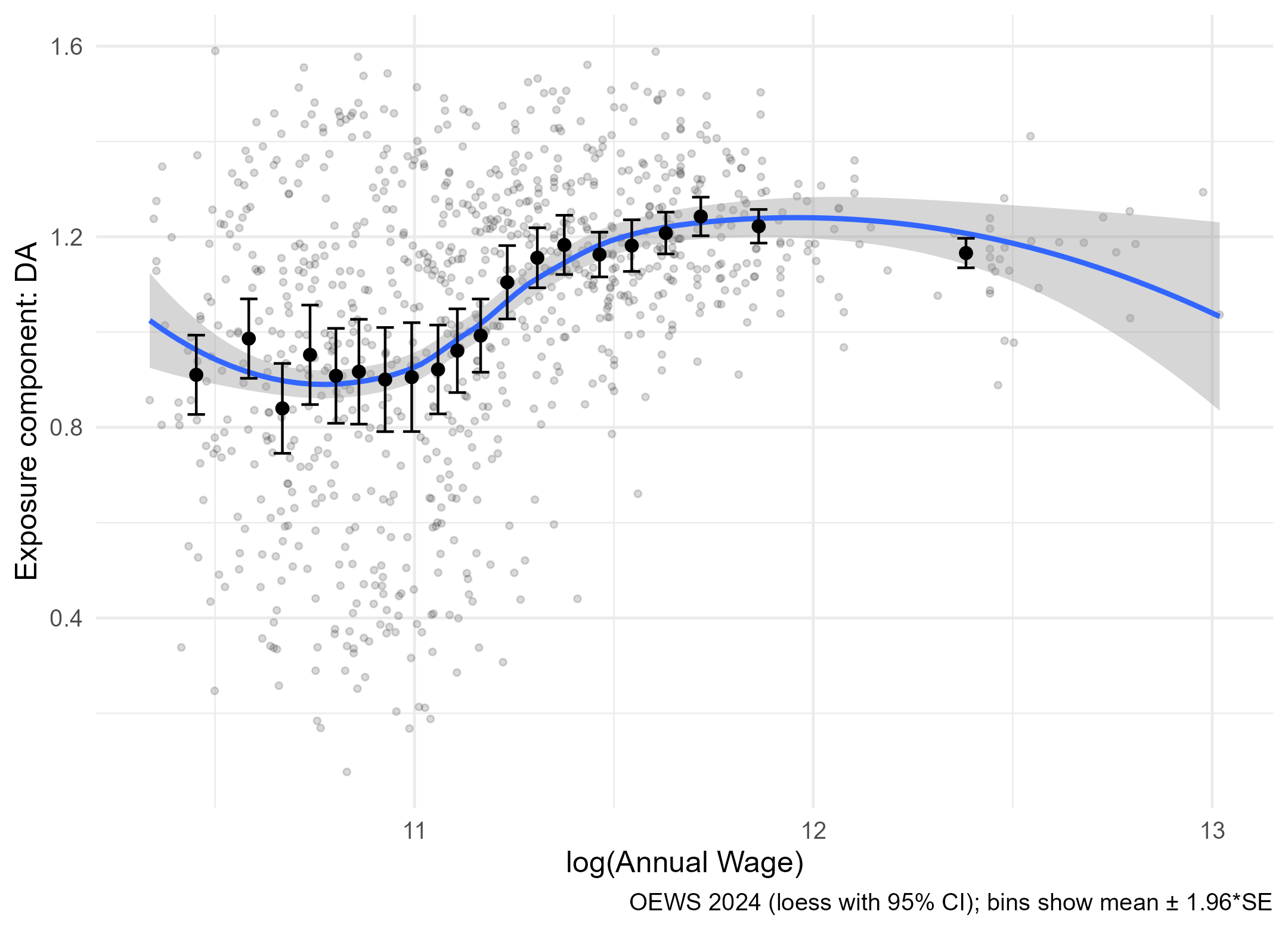}
  \caption{ Data Abundance (DA) exposure vs log wages, 2024}
  \label{fig:subhyp2}
\end{figure}

\begin{figure}[h]
  \centering
  \includegraphics[width=0.7\textwidth]{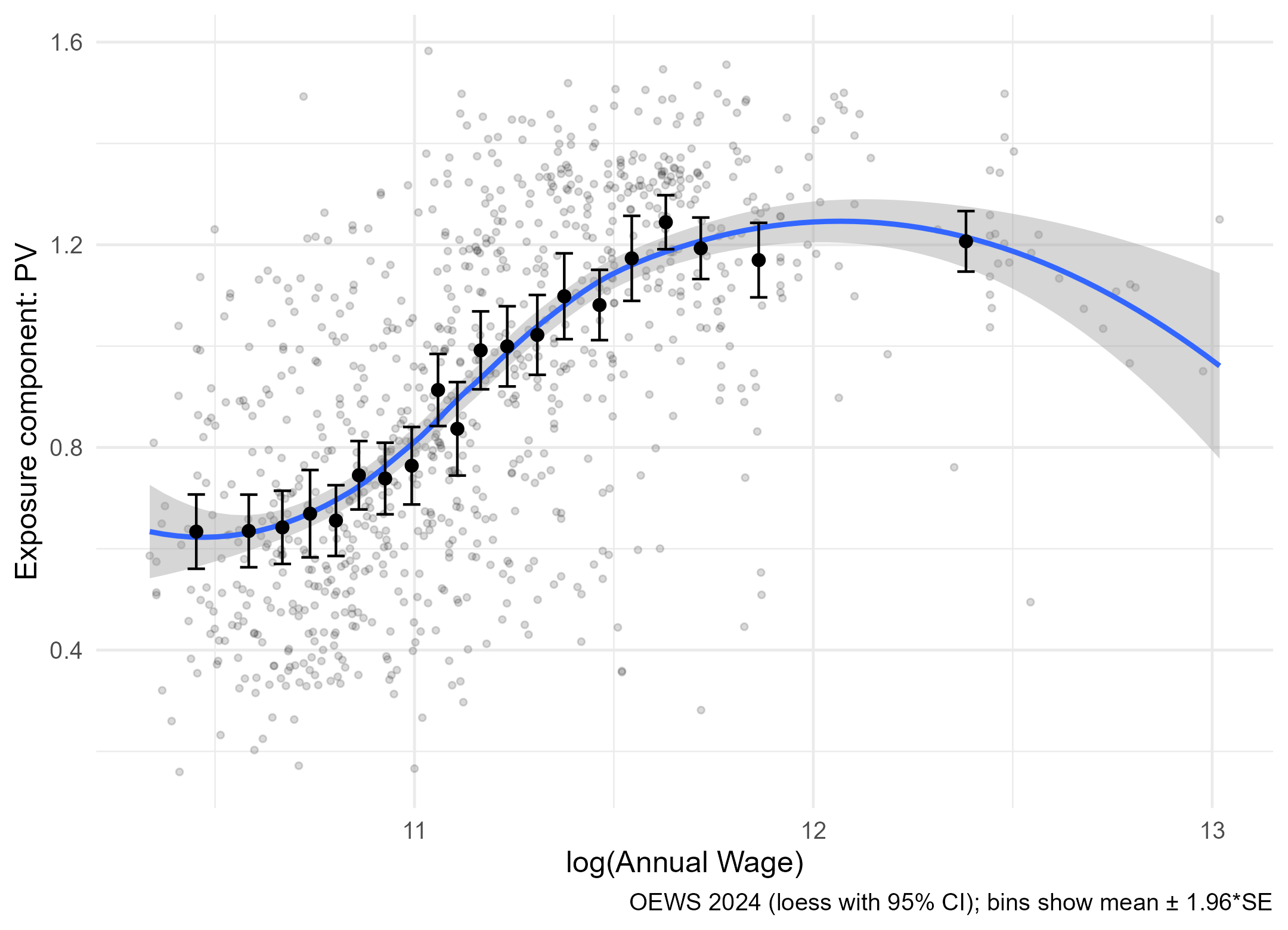}
  \caption{Productivity Variance (PV) exposure vs log wages, 2024}
  \label{fig:subhyp3}
\end{figure}

\begin{figure}[h]
  \centering
  \includegraphics[width=0.7\textwidth]{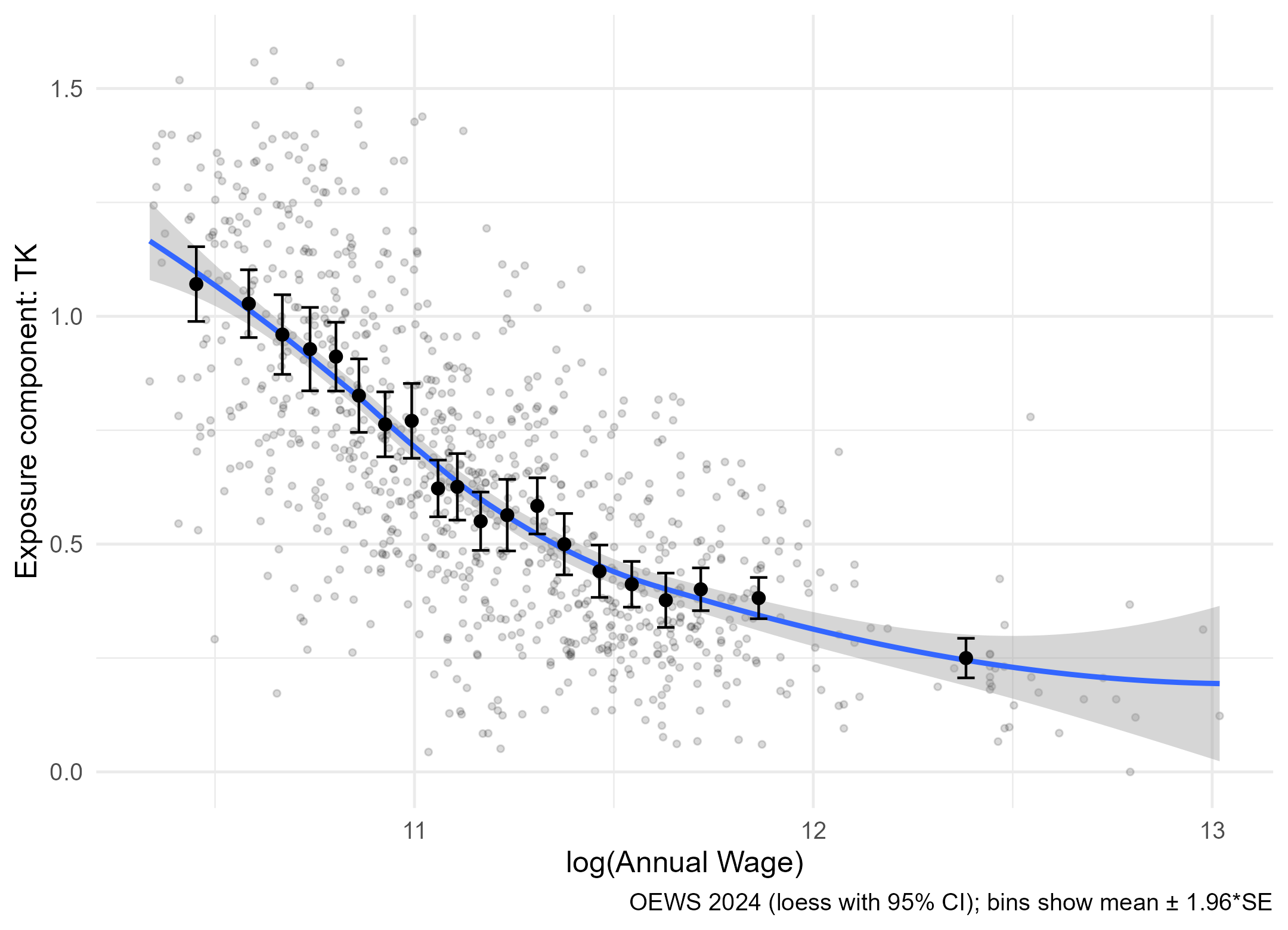}
  \caption{Tacit Knowledge (TK) exposure vs log wages, 2024}
  \label{fig:subhyp4}
\end{figure}


\clearpage

\newpage

\RaggedRight

\bibliographystyle{dcu}
\bibliography{bib}

\end{document}